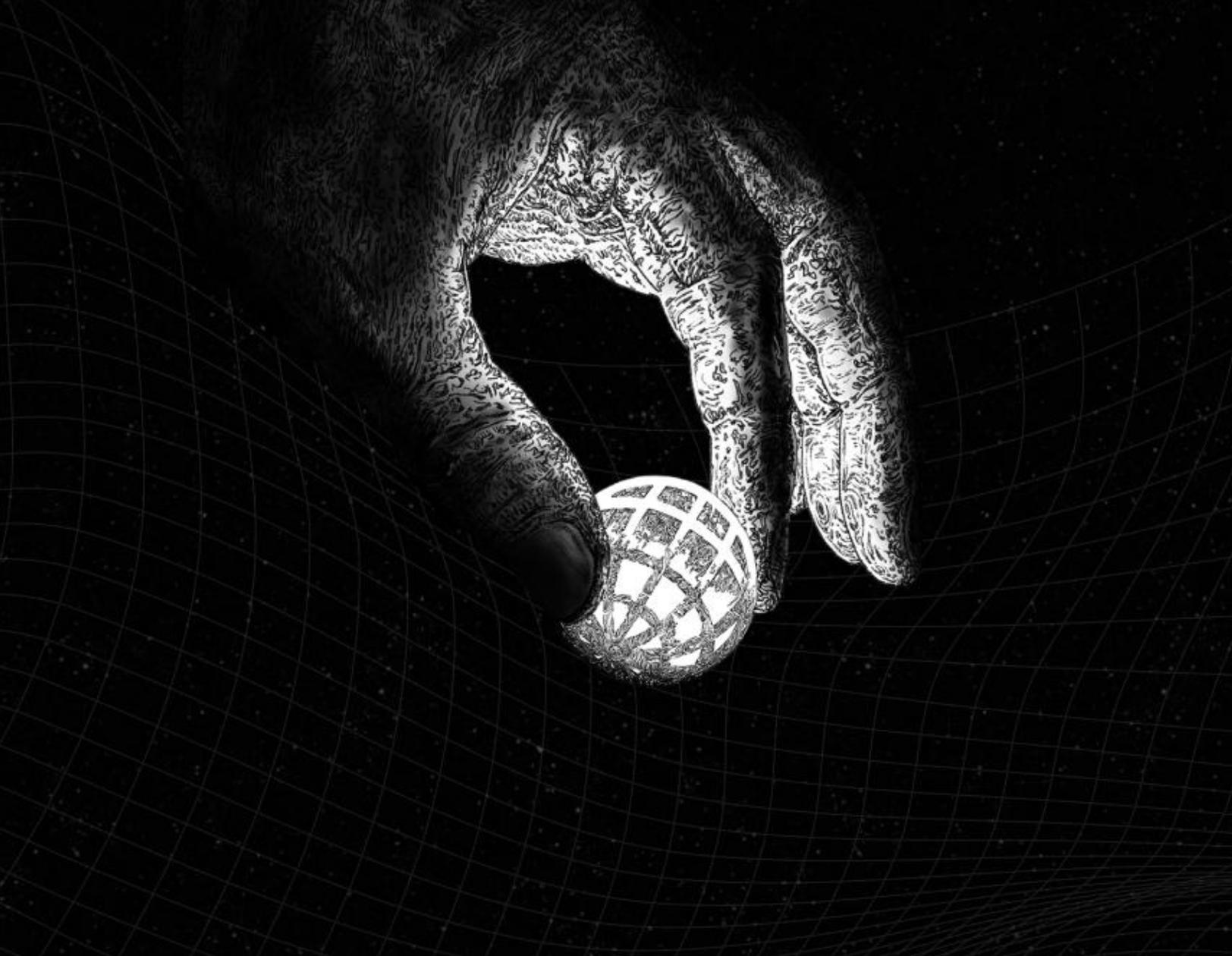

# Social Dynamics of DAOs:
## Power, Onboarding, and Inclusivity

DAOstar, July 2025

DAO*

# Executive Summary

This report explores the often-overlooked cultural and social dynamics shaping participation and power in DAOs. Drawing on qualitative interviews and ethnographic observations, it shows how factors such as financial privilege, informal gatekeeping, visibility bias, and onboarding structures create barriers to meaningful inclusion. While DAOs are frequently framed as permissionless and egalitarian, the lived experiences of contributors reveal a more complex reality, one in which soft power and implicit norms determine people's position within DAOs.

Instead of offering solutionist prescriptions, this report argues for a deeper cultural reflection within the DAO ecosystem. It highlights that decentralisation is not solely a protocol-level feature, but an ongoing social process that requires intentional cultivation of trust, belonging, and epistemic plurality. With this report, we want to sharpen the collective awareness of structural blind spots and call for building more inclusive and culturally conscious decentralised systems.

# About the authors

**Victoria Kozlova** is a research fellow at DAOstar and co-founder of acurraent. A multimedia designer by day and interdisciplinary researcher by night, her research interests span sociolinguistics, feminisms, digital colonialism, and user experience.

**Ben Biedermann** is a research fellow at DAOStar and the Resident Protocol Product Lead at Nouns Builder. He also pursues a PhD at the Islands and Small State Institute at the University of Malta and co-founded acurraent.

This report is a publication of DAOstar (or DAO*), the standards body of the DAO ecosystem.

# Contents

# 1. Introduction

Decentralised Autonomous Organisations (DAOs) have been discursively positioned as a breakthrough in collective coordination, for they are transparent, permissionless, and community-governed, they promise to flatten hierarchies, expand access, and reshape how humans collectively govern shared resources. Given the combination of software logic and ideals of egalitarian participation, DAOs indeed put forth a compelling experiment for a more decentralised future of collaboration.

Yet, this experiment has given way to disillusionment. A growing number of researchers and practitioners acknowledge that in practice many DAOs are far from the decentralised ideals. Discussions on why DAOs have failed largely point to governance apathy, token concentration, centralised decision-making, and misaligned incentives.

However, far less attention has been given to the cultural, social, and relational dimensions of decentralised governance, specifically, to the lived experiences of those taking part in these systems on a daily basis. How do contributors find their way into a DAO? What forms of social capital shape legitimacy? Who gets recognised, and who remains peripheral despite participating? This research project emerged from the conviction that decentralisation without cultural plurality and epistemic inclusion cannot be considered decentralisation, rather coordination within a closed social loop.

While decision-making power may be formally distributed through tokens and roles assigned to contributors, influence in social contexts often circulates through less visible channels. These include, for instance, shared assumptions between certain individuals, dominant communicative styles that are more favorable than others, and access to specific prerequisite technical knowledge. Hence, even DAOs that appear decentralised on the protocol level, may still reproduce cultural homogeneity and soft exclusion in practice. As a result, decentralisation becomes a structural shell, a discursive marker that neglects how power is concentrated beyond the surface.

Drawing on qualitative research and ethnographic observations, this project makes inquiry into *how power dynamics operate beyond the visible layers of DAO coordination*. The analysis is grounded in technofeminism, which offers approaches for understanding how social imbalance and systemic exclusion shape experiences in digital systems. It also leans on a phenomenological orientation, attending to how contributors feel and make sense of their DAO involvement. Hence, this is not a study of governance mechanics, but of *governance as lived reality*, shaped by whose voices get heard, and how contributors position themselves within spaces that are open in theory, but culturally gated in practice.

A common expectation placed on researchers conducting critical sociocultural inquiries into digital systems is to offer not only critique, but also clear frameworks or toolkits to fix it. In other words, raising concerns is often met with "So what's your solution?", as though the validity of critique hinges on its ability to produce "solutions." In Web3 and broader tech culture, this demand often takes the form of solutionist urgency, which essentially results in belief that every issue can be patched with code, upgraded governance, or a new dApp. When presented at a conference, this research was met with the same refrain, hence, the presence of this paragraph.

For those seeking answers, this paper does not offer a fix. It resists the impulse to find a solution. Instead, the aim of this inquiry is to unsettle the illusion of coherence, what Mohanty (2003, qtd. in Ahmed, 2013) calls *harmonious empty pluralism,* that dominates the DAOscape. This work invites a collective reflection, rather than offering closure, by bringing up the informal, cultural, and social dynamics often left unexamined in the tech-heavy spaces. The aim is to sharpen the lens through which we view decentralised participation and to challenge the idea that technical (or economic) infrastructural decentralisation alone is sufficient for inclusion.

# 2. Decentralisation Meets Reality

DAOs emerged in the Web3 discourse with a grand promise to reinvent how people coordinate, govern and work together. The ethos of *DAOism* (Dylan-Ennis & Kavanagh, 2024), "involves a commitment to decentralization, [and] skepticism towards hierarchies". In theory, a DAO distributes authority among its members instead of concentrating power at the top, as is the case in most traditional organisations with hierarchical structures. DAO proponents believe this model can strengthen the functioning of distributed online communities by relying on "wisdom of the crowds" (Buterin qtd. in Raj, 2022) instead of centralized control. The concept of DAOs, in terms of its features and characteristics, is still a contested concept (Hassan & De Filippi, 2021), and we are not going to dive deep into the conceptual frameworks here. There is an expanding body of literature on DAOs that can be read on this topic (DuPont, 2017; Hassan & De Filippi, 2021; Kerckhoven & Chohan, 2024; Owocki, 2025; Sharma et al., 2023). Instead, we will focus on how they are positioned and the problems they are claimed to address. In essence, DAOs are envisioned as digital-native, leaderless organisations where communities vote on proposals (often using governance tokens), which is supposed to ensure that no single person has unilateral control.

For instance, longstanding industry members like Kevin Owocki (2025) position DAOs not as "just better vehicles for innovation, coordination, collaboration, and wealth creation", but as structures that can help "make the global economy fairer, more inclusive, and more democratic". In summary, the promise of DAOs put forth by Web3 practitioners include: (a) more democratic governance (due to a broader base of token holders compared to a board of directors), (b) community ownership of assets and outcomes, (c) greater transparency (since all transactions are recorded on-chain), and (d) the ability to coordinate at scale globally without traditional intermediaries.

However, a gap between the promise of DAOs and on-the-ground reality has appeared as DAOs have been moving from these ideals to implementation. Both academic and grey literature point out that many DAOs are not as "flat" or autonomous as people would like them to be (Austgen et al. 2023, Tan et al., 2023, Tai, 2022). The discourse on why DAOs have failed has also been present for some time now, especially around "Crypto Twitter" (X). Indeed, it is common for DAO votes to be decided by a small number of "whales," a handful of addresses holding the majority of tokens, or delegates.

These are usually legitimised by another problem in DAOs – governance fatigue. Arguably, not many people have the interest or capacity to monitor the forums all the time and familiarise themselves with open questions. Ironically, some degree of delegation or representation, while seemingly centralising, may in fact improve decentralisation by mobilising the silent majority, which is why the delegation mechanisms have been in place in DAOs. Especially the largest DAOs, such as Arbitrum, Optimism, and Maker, make use of delegation structures. Although the delegated DAO model experiences increasing popularity, it is not always adopted for improving democratic principles or the ideal of decentralisation. Instead, token delegation allows stakeholders with vested interest in the success of the project to exert influence without risking facing scrutiny by regulators for the apparent lack of decentralisation. The delegate model, therefore, runs the risk of reinforcing hegemonic structures of decision making and discouraging open engagement.

At present, it looks like not only the "D", but also the "A" of DAOs is flawed. Although the "A" in DAOs stands for *autonomous*, where, in their utopian form, human involvement and decision-making are pushed to the edges, this is not a universally accepted reality but rather a vision outlined in the early days of DAOs (Buterin, 2014, summarised by Merk, 2024). In such visions, smart contracts and other automated systems, e.g. AI agents would coordinate action and enforce rules with minimal human input. But an organisation is impossible without humans. Human interaction still cannot be avoided, even in DAOs that strive for a maximum of automation in governance (Dylan-Ennis, 2024, p. 43). Once humans are involved, attention should also be placed on human behaviours, social relationships, and power dynamics. These factors likely cannot (and probably should not) be sidestepped by introducing a protocol to obscure the social layer.

Thus, beyond token metrics, it is important to pay attention to the sociocultural power dynamics that lead to power imbalances and might cause some of the issues in DAOs, voter apathy and governance fatigue included. Primavera De Filippi and Marc Santolini (2022) introduce the notion of "extitutions" to capture the interpersonal relationships and social dynamics that exist in organisations alongside formal rules. They are also present in any DAO, alongside the on-chain governance rules which can be considered the formal "institutional" layer of coded voting rights and contributors' permissions (De Filippi & Merk, 2024). In DAOs, the *extitutional* layer manifests through social interactions such as reputation, friendships, informal norms, and off-chain coordination. These factors can and in fact do recentralise power in subtle, and sometimes not-so-subtle ways. For instance, core developers or founders might retain outsized social influence simply due to their expertise or charismatic leader-like position, even if they hold modest token shares. Informal "cliques" of insiders can coordinate votes off-chain or sway community sentiment in forums simply due to their higher reputation within the community. As Dylan-Ennis (2024) puts it, alongside the technical and economic components – 'hash' and 'cash' – a significant part of Web3 communities is also 'bash', the social layer of DAO operations, mostly happening on Discord, Telegram or DAO-specific forums.

And the 'bash' is just as important, or maybe even more important, than 'hash' and 'cash' for a sustainable community building. As put by De Filippi & Merk (2024):

> The extitutional culture is simultaneously shaped by, and a driver of the interactions, social relationships, and shared experiences of individuals within the organization. It informs behaviors, socialization processes, and how individuals interpret and respond to the formal institutional structure."

Building on the notion that a cohesive extitutional culture strengthens member alignment, participation, and trust (De Filippi & Merk, 2024), this research focuses on uncovering the social dynamics that might stand in the way of increased sense of belonging within DAOs nowadays.

The dissonance between DAO ideals and practice (captured in Figure 1) has parallels in organisational theory at large. Scholars of organisational sociology have long noted that the formal flattening of hierarchy does not necessarily result in equal participation (Crozier, 2010). Rather, informal hierarchies and status dynamics naturally emerge in any organisation, often along lines of uneven distributions of cultural and social capital (Bourdieu, 1986) among contributors. Feminist scholarship, in particular, offers insights into how the narratives around valuing diversity, or decentralisation, in the case of DAOs, can become "performative" gestures that signal virtue without producing real change. Sara Ahmed (2012), for example, critiques traditional organisations, particularly universities, for institutionalising diversity in a way that protects themselves without adequately recognising or embracing diversity. Transposing this to Web3, we might wonder whether the promise of decentralisation is just a brand slogan that enables organisations to operate with business-as-usual power structures. This perspective should urge us to critically examine who actually holds power and benefits in these new decentralised systems.

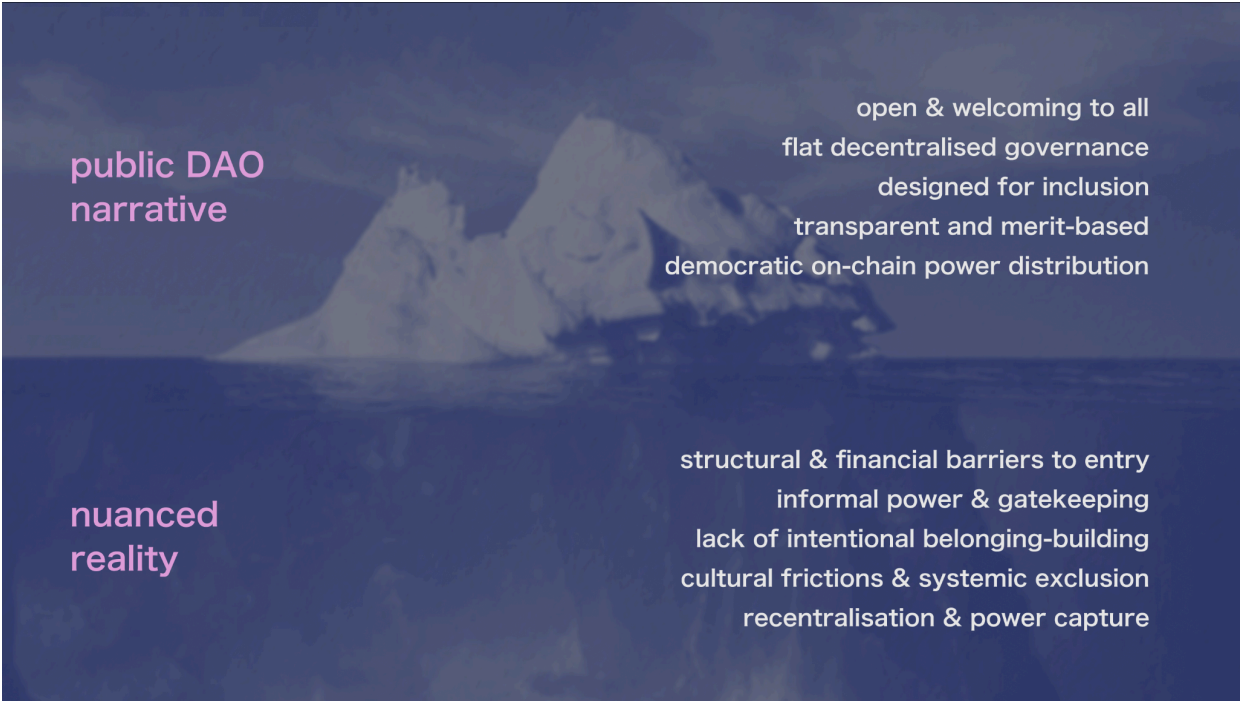

*Fig. 1. A clash between DAO ideals and current practice*

Feminist scholars, like Judy Wajcman (2004) also remind us that technology alone cannot and will never erase social biases. Instead, it is very well equipped to reproduce them on an even larger scale. Wajcman's concept of *technofeminism* highlights the mutual shaping of technology and society and reminds us that social values must be intentionally built into tech communities.

In conclusion, the literature paints a picture of "decentralization meets reality" as a mix of inspiring potential and sobering challenges that are yet to be solved. DAOs do offer a novel toolkit for organizational experimentation. Moreover, some have successfully funded projects, coordinated global communities, and demonstrated more transparent governance than comparable traditional corporations. However, critical inquiry suggests that decentralization is not a one-time achievement but a continuous process of design and community cultivation. Truly delivering on the promise of DAOs and cultivating durable communities will likely require blending technical fixes with deep cultural work, for example, encouraging diverse voices, practicing inclusion, and developing norms that resist social recentralization.

# 3. Studying DAO Culture from the Inside

The project was initially framed as an inquiry into the perceived cultural homogeneity among DAO contributors. The research idea was based on ethnographic observation and inspired by ongoing discussions in diversity-focused communities and blog posts. While conducting the first couple of interviews, however, the research organically evolved from asking structured questions to listen to people speaking candidly, often not without frustration, about the cultural and social obstacles they encountered in their DAO engagements. As such, the research took shape as a form of experience-led inquiry to capture lived stories from within the DAOscape and group them into emergent thematic groups. This transition was welcomed because it not only reflected the participants' energy and needs but also aligned with an inductive approach that prioritises meaning-making from the ground up.

We conducted ten direct semi-guided qualitative interviews and engaged in informal conversations with multiple respondents throughout and prior to this research project. The interviews ranged from 30 to 60 minutes, and most were automatically transcribed with the informed consent of the interviewees.

After the interviews, we carried out an inductive thematic analysis, manually identifying recurring patterns and narratives across the dataset. We annotated the transcripts and memos to allow themes to emerge through repeated engagement with the data. Through this process, we ended up with six distinct broader themes that are discussed in detail in the following section.

Our interviewees were active in or had visibility of the inner workings of the following ecosystems: Arbitrum DAO, Nouns DAO and Noun ish DAOs, DAOHaus, RaidGuild, SEED Club, Scroll DAO, Asterisk DAO. Notably, most of our participants were working in non-technical roles, as researchers, marketers, governance analysts, designers, and community managers. As such, the findings presented here primarily reflect the experiences of non-technical contributors, whose perspectives often go underrepresented in technical spaces. This may be a strength, given the widespread dominance of engineering voices, but it also means that the findings are not representative of all contributor types.

Most interviews and informal conversations for this research were conducted by the female co-author and the majority of contributors we engaged with identified as non-male. Given the limited number of interviews, contributor roles, and gender representation, we acknowledge the limitation of this research in not being representative for the entire DAOscape. Nevertheless, we consider the issues raised by the interviewees highly relevant for anyone committed to building truly decentralised cultures.

# 4. Frictions in Participation: Themes from the Field

In this section, we outline the thematic groups that have emerged throughout our conversations and analysis of the transcripts. The themes are as follows: (1) Financial Filters and Infrastructure Gaps, (2) Who Holds Power? Gatekeeping and Soft Control, (3) Reputation Systems and Visibility Bias, (4) Creating and Performing Belonging, (5) Cultural Fit and the Myth of Neutrality, and (6) Soft Power as Acts of Resistance. These themes are not mutually exclusive, rather, deeply entangled and reinforce each other in subtle ways.

## 4.1 Financial Filters and Infrastructure Gaps

While DAOs are often described as permissionless spaces, participation in practice is frequently shaped by a range of structural and economic constraints. It is acknowledged that participation in Web3 *per se* is a privilege. The financial burden of this participation is probably one of the most obvious barriers to entry, notably, for both people outside and within the industry.

One of the reasons DAOs rely on blockchain is to use smart contracts to coordinate social or economic interactions (Wright & Law, 2021). As such, DAOs typically operate with a prominent financial component, usually in the form of a shared pool of funds, also called treasury, collectively managed by members (Dylan-Ennis, 2024, p. 43).

To begin with, let us return to the promise of global inclusion and collaboration without borders, or the idea that DAOs don't care where one lives. While digital collaboration may indeed be made easier through blockchain-based tooling, it would be a fallacy to declare DAOs are truly open and accessible to all, given the systemic inequalities in place. In reality, DAO participation, although attracting people from across the globe, remains largely limited to the most privileged percentage of the world population, both financially and socially. This is not only evident in the distribution of tokens but also in the politics of meeting times, which usually favour American timezones over Eastern Europe, the Middle East, and Asia.

Even among those who have made their way in, or are in the process of joining, DAO participation is often shaped by structural and economic concerns. Entry into a DAO usually requires possession of governance tokens, the acquisition of which is not accessible to anyone. The cost of shares varies depending on a DAO, but rarely starts below 100 USD. Several interviewees mentioned they simply could not afford to purchase an NFT representing membership, or enough tokens to meaningfully participate in the early stages of their DAO journeys, for instance, in the Nouns community. In this way, participation becomes a matter of capital access.

On top of that, wealth-based influence was also mentioned by several interviewees. In token-based voting, large token holders are able to strategically delay votes, exercising power not just through accumulation but through control over timing and outcome.

Alongside financial barriers to entry, many contributors, particularly newcomers to a DAO, are expected to engage in weeks or months of unpaid labour before receiving any recognition or compensation. This procedure is often normalised as part of Web3's "prove yourself" culture. Needless to say, this model systematically advantages those with the means to work without immediate pay, and more often than not excludes those without such an option.

Particularly notable is the type of labour that remains un-/underpaid or underrecognised most frequently. Tasks like moderation, onboarding, documentation, conflict resolution are important for a functioning of any collective, a DAO included. Yet, these roles are rarely reflected in contributor roles or compensation systems. On one hand, it is understandable, as DAOs rely on people showing up and staying engaged because they believe in the long-term vision of the project. On the other hand, when certain tasks and responsibilities are systematically undervalued, often justified by ideals of volunteerism or decentralised ethos, it creates a barrier to sustained engagement for anyone without a financial safety net.

At the same time, there was a concern raised that purely financial incentives can also attract opportunistic contributors and risk undermining long-term commitment and community-building values. In this sense, the question is not only whether the work is paid, but how value is recognised and whether compensation structures are designed to support sustainable participation, rather than short-term transactional engagement.

Finally, participation in DAO governance assumes much more than basic, i.e. uncensored internet access from a location that is not blocked due to concerns over Office of Foreign Asset Control (OFAC) regulations, as is often claimed when talking about permissionless engagement. It presumes a degree of fluency or at least prior exposure to governance forums, proposal platforms, wallet mechanics, and Discord's *affordances* (understood here as platform-specific action possibilities; Bucher & Helmond, 2018). Not to mention the cultural tone and pace of Web3 interactions. As one interviewee pointed out: *"It is not that the information is not out there or kept secret. It is just scattered and nobody tells you what to look for"*.

On top of this, staying active in DAOs requires temporal availability to read proposals, keep up with threads, attend community calls, and react to pings. As put by one of the interviewees, *"keeping up with all the conversations [is] almost like a full-time job."* Thus, for contributors with care responsibilities, full time jobs, or poor connectivity, staying present is structurally more complicated. Over time, this creates a loop of exclusion, where members with more bandwidth, flexibility, social and financial capital dominate the conversation.

## 4.2 Who Holds Power? Gatekeeping and Soft Control

The concentration of influence is not merely a structural artifact, but the result of complex social dynamics, informal gatekeeping, often unintentional, and the soft exercise of control. As in other online communities (Gasparini, Clarisó, Brambilla, & Cabot, 2020 ctd. in de Filippi & Merk, 2024), in DAOs, a small core group of people tends to drive the majority of contributions and decision-making. This is a well-documented phenomenon and, to some extent, an inevitable outcome of collective human behaviour.

One notable mechanism of influence in DAOs, as observed by interviewees, is *pre-lobbying*, the informal shaping of proposals before they even reach formal governance stages. Contributors with good standing in the community, close ties to core members, or strong reputational capital often have disproportionate sway over what is considered "vote-worthy". Proposals frequently go through several iterations to align with the expectations of influential actors. In such contexts, founder (or OG) charisma, insider networks, and perceived loyalty regularly outweigh token-based voting power or public contribution history. Interviewees reported feeling discouraged from proposing new ideas altogether, when they assumed a key core contributor is unlikely to support them and others may follow suit or remain silent out of fear of losing face. This dynamic extends to uneven accountability structures. As one contributor remarked, *"Who gets held accountable and who doesn't is actually very political… if you have influence and funding, people sometimes won't say anything."*

This environment cultivates what can be described as soft control. Raising sensitive topics such as diversity, equity, and inclusion often follows the same path as supporting an unpopular proposal: it may carry reputational costs. As a result, contributors may self-censor or avoid engaging with necessary but uncomfortable issues.

Turning back to the concept of extitutional culture (De Filippi & Merk, 2024), defined as the fluid, interpersonal logic that operates outside formal structures but still profoundly shapes them, we can see how these dynamics manifest in DAO governance. This culture "informs behaviours, socialization processes, and how individuals interpret and respond to the formal institutional structure" (De Filippi & Merk, 2024). On the example of Moloch DAO, De Filippi and Merk (2024) describe how a single contributor bribing others can distort the functioning of a DAO. Such influence does not always take the form of explicit bribery. It can be exercised through social norms, implicit expectations, reputational alliances, or the subtle pressure to conform to the dominant sentiment. Contributors may support proposals not purely on its potential, but out of loyalty, obligation, or fear of losing face.

Core contributors often do not intend to exclude others, but informal scrutiny and aversion to reputational or financial risk can lead to the quiet sidelining of newer or less connected participants. In practice, DAO governance frequently depends less on token-weighted votes than on who is heard, who is trusted, and who feels safe enough to speak.

## 4.3 Reputation Systems and Visibility Bias

Reputation systems are often positioned as a potential solution to overcome the shortcoming of token-based governance. It is a way to measure contributions beyond financial wealth, and reward merit over speculation. Several Web3 projects have experimented with alternative reputation systems.

Among these, tools like Coordinape and Gitcoin Passport that reward peer recognition or aggregate identity data. Projects like Karma GAP and Wonderverse took a different approach by tracking visible contributions such as voting or milestones and task completion to hold projects accountable and track impact. Others, like Otterspace and Hats Protocol, focused on roles and responsibilities by issuing metadata-rich credentials or badges on-chain, which may have permissions attached. In practice, these systems tend to privilege a narrow scope of activity, particularly, those that are public, visible and easily measurable. This leaves social background, relational and emotional labour largely unrecognised. Moreover, reputation and ranking systems do not affect all contributors equally. For some, they can serve as motivation; for others, they are demotivating or even alienating (Cook et al., 2009 ctd. in De Filippi & Merk, 2024). An alternative approach is the Tokens of Appreciation Protocol (TOAP), which introduces peer-to-peer recognition through flexible programmable tokens. Instead of tracking performance or output, TOAP emphasises relational value and allows contributors to acknowledge meaningful support or presence that often goes unrecognised in traditional reputation systems.

Across our interviews, contributors described how presence, posting frequency and fluency in Web3 discourse become *de facto* tokens of recognition. DAOs tend to reward and value more contributions by those who speak loud and often, actively engage in conversations regardless of the input quality they produce, and express themselves confidently, especially in English. Contributors who can react and respond quickly to ongoing discussions and who regularly appear in calls are more likely to be considered "active" or "valuable."

As one participant put it: *"If you're not loud, it's like you're not even there."* This dynamic creates a skewed form of visibility, namely, one that equates merit with performance, and responsiveness with leadership. This reinforces performative participation where being seen is valued more than what is actually contributed. As reported by some of our interviewees, this tendency sidelines those who prefer quiet, behind-the-scenes work, whether that means maintaining infrastructure, writing, or contributing through less conversational forms of labour. These contributions are often foundational to DAOs' operational backbones but rarely recognised, especially when they are not accompanied by active public commentary or self-promotion.

## 4.4 Creating and Performing Belonging

"DAOs connect people together through blockchain-based protocols and code-based systems, focusing on achieving a shared social or economic mission" (Wright and Law, 2021, p.5). Yet, despite this mission-oriented framing, we have observed a notable absence of intentional belonging-building in most DAOs. Open access does not automatically translate into a sense of inclusion and recognition.

Drawing from the work of social anthropologist Pfaff-Czarnecka (2011), *belonging* can be understood not simply as identification with a group, but as an emotionally charged social location of oneself rooted in commonality, mutuality, and attachment. It is something people feel, not something that can be handed to them. The experiences making one feel belonging are inherently social, reciprocal and often informal. In contrast to the more categorical concept of *identity*, which defines boundaries against others, belonging is about widening borders, maintaining social ties, and participating in creation of shared experiences (Pfaff-Czarnecka, 2011). While DAOs may be structured around shared goals or a specific product, they rarely foster this kind of social and emotional embeddedness.

Belonging emerges naturally but it can also be cultivated. While it cannot be manufactured on demand through onboarding checklists or formal frameworks, belonging can be intentionally enabled by creating the right conditions. Belonging can be nurtured through such extitutional factors like sustained human engagement, recognising the (non-financial) value of individual contributors (Rashid et al., 2006 ctd. in De Filippi & Merk, 2024), cultural openness, and emotionally generous feedback.

Our interviewees reported absence of this kind of culture in many DAOs, which is notably more prominent in larger communities and more product-oriented DAOs. The absence of such intentional scaffolding means that belonging is left to chance and usually only remains accessible to those who already align culturally with the dominant group. Contributors might leave not only because of technical failures or discontinued interest, but also because no one made space for them to feel anchored.

On the technical level, the affordances of blockchain make it possible for certain structures of belonging to be computationally enforced through tokenisation, smart contracts or DAO voting protocols. Yet, as Zeilinger (2024) argues, there is an opportunity here to reimagine blockchains "not as property-oriented enclosure systems, but instead as structures of belonging." These systems can support relational presence and collective stewardship, but only if they are socially enacted.

For many people interested in joining a DAO, the difficulties start at the stage of onboarding. Most DAOs still rely on text-heavy, documentation-based onboarding processes, which automatically presumes contextual awareness and prior Web3 knowledge. There is probably no better way to make someone feel disoriented than to throw them to the Docs to figure it out by themselves. While informal efforts do exist, they often rely on social proximity, i.e. a friend helping a friend with onboarding. This makes access to DAOs uneven and unsustainable through onboarding people from the same circle and creating a homogenous culture from the start.

More structured approaches, like cohort-based entry or one-on-one mentorship, were highlighted as more supportive by our interviewees. For example, one interviewee described an instance where two new contributors joined Little Nouns for the first time and received detailed feedback on their proposal:

> "Everybody gave feedback… they were super nice. It was very refreshing to see people take two or three hours out of their day to help carve out the proposal and explain what the DAO actually looks for."

While there was no official onboarding channel, the interaction resembled a form of onboarding through sustained presence and care.

RaidGuild does a great job of the initial onboarding via cohorts, where a lot of explanation of the inner workings and handholding is happening. Having someone explain the Docs to you, meeting fellow community members, and getting a feel of how it feels to be part of this DAO makes people feel seen, guided, and gradually integrated.

While other DAOs, as reported by some of our interviewees, create a sense of excitement for newcomers who made their way past the Docs in the beginning, this excitement and support often fades quickly, creating a "leaky pipeline" where early engagement vanishes without emotional anchoring. We borrow the term *leaky pipeline* from STEM fields, where it describes systemic attrition, particularly among women and underrepresented groups, not due to lack of interest or ability, but because the surrounding environment fails to support their sustained participation.

Mentorship also emerged as a distinct theme to sustainably retain and motivate newcomers. Several interviewees described that receiving personal guidance, encouragement, or even a simple invitation to contribute somewhere their skill might be valuable, significantly increased their confidence and long-term motivation. Perhaps it is time to move beyond the individualistic ethos of "just show up and figure it out," and instead cultivate a culture of reaching out, offering guidance, and sharing responsibility.

Several contributors from larger DAOs described feeling lost or disconnected in the absence of substructures — such as micro-communities, subDAOs, guilds, or focused working groups. From both interviews and observations, it became clear that the smaller the scale, the more space there is for relational bonding and creative participation. We particularly liked the framing of one of our interviewees, Sarah Smith from DeepWork, about creating *microcosms*, small, self-contained environments that can serve as safe zones for experimentation without risking disruption to broader DAO processes. These microcosms could allow for the low-risk testing of concepts, with the potential to scale if something proves to be effective, but also help cultivate the sense of belonging: *"I think that's really important is feeling like you're part of the community, but you also have a specific community inside of it that has a purpose too."*

Lastly, feeling valued, as many participants emphasised, does not come from titles or token holdings, but from small interpersonal cues: someone remembering a past contribution or a personal difficulty, a compliment on work, a follow-up message. These "micro-acts of recognition" are often overlooked in digitally mediated settings, yet they are the acts creating social belonging.

This pattern reflects what sociolinguist Jan Blommaert (2018) describes as *light groups*. They are translocal, internet-native formations built around knowledge exchange rather than durable social ties. DAOs, much like other digitally mediated communities often operate as communities of knowledge, where participation depends on navigating complex information flows and informal hierarchies between 'experts' and newcomers (Blommaert, 2018). While this structure may offer broad access, it rarely provides room for building relations or sustained support that is needed for creating a strong sense of belonging. As a result, membership in such communities is characterised by fluidity and temporality.

## 4.5 Cultural Fit and the Myth of Neutrality

While often framed in terms of protocols and coordination, DAOs are, fundamentally, cultural formations embedded within a broader ethos of decentralisation. The very idea of a decentralised autonomous organization is shaped by longstanding ideals around resistance to central authority and open-source collaboration. First and foremost, these ideological roots have laid out a foundation to technical experimentation, but as cultural formations they also reflect the worldviews of the early adopters and builders of the space: libertarian-leaning, tech-optimistic and deeply embedded in the crypto-native internet culture. This inevitably shapes the design and operation of DAOs, i.e. who joins them, how people interact, and what is considered valuable, just as much as the code itself.

DAOs vary widely in their purpose and tone. Some are product-oriented and focused on execution; others lean more into community, art, culture, or specific social causes. The differences between them are not merely functional, they reflect deeper cultural orientation and the internal norms of each DAO. Several interviewees noted a clear divide between DAOs driven by "fun and culture", such as PizzaDAO, and those more focused on financial control or governance optimization in their daily operations. The latter, particularly when dominated by VC stakeholders, tend to privilege professionalism, technical language, and efficiency. Unfortunately, this often occurs at the cost of inclusivity or cultural openness, making participation feel more exclusive compared to their "fun and culture" counterparts.

As part of broader Web3 culture, DAO spaces remain male-dominated and frequently reflect male-coded norms, especially DAOs focused on decentralised finance. These norms include competitiveness, hyper-rationality, and status performance through visibility. While these traits are often viewed as "neutral" or even desirable in order to succeed, they are in fact situated cultural expressions, which have the potential to exclude those who do not share that background. The historical underrepresentation of women and marginalized groups in tech has not only affected who participates, but also how technological spaces are designed and experienced today (Wajcman, 2010).

One interviewed contributor, despite her early and continuous engagement across multiple DAO communities, described feeling especially unwelcome in VC-heavy DAO spaces. She mentioned experiences of casual racism, classism, and unaddressed sexism, often "just floating around", sometimes framed as jokes, other times more ambiguous in tone. Her experience is not isolated and reflects a broader pattern of exclusion despite contribution. While not directly excluded, being a minority in a space where minorities are joked about is clearly far from a welcoming experience.

Soft barriers to entry and retention in DAOs often arise through in-group dynamics and shared cultural references that remain inaccessible to newcomers unless someone takes the time to explain it to them. These references can include past events, niche memes, or unwritten social codes, such as expectations around proposal tone and structure, communications etiquette, all of which subtly signal alignment and belonging. As one interviewee shared, one must first invest significant effort in understanding these cultural expectations and researching how to proceed: *"Otherwise, your contribution might get dismissed or rejected just because they don't fit with the social norms established."*

The widespread use of English as *lingua franca* and reliance on Western time zones for community calls further marginalise contributors from other linguistic and geographic contexts. Until these individuals acquire enough standing within the DAO to request more inclusive scheduling, they are effectively left to adapt. Language barriers are even more complex to address, though, admittedly, in this iteration of the research, we only focused on English-speaking DAOs and contributors, which limits our ability to generalise about linguistic diversity across the ecosystem.

To sum up, we could argue that exclusion is not just about identity, it lies at the intersection of technological and cultural infrastructures. What we mean by that is that the lack of attention to onboarding practices, mentorship, or accessible language reflects a culture that assumes everyone can and should figure it out *alone*. As Wajcman (2010) and Haraway (1988) remind us, technologies and systems reflect the perspectives and values of those who create them.

## 4.6 Soft Power as Acts of Resistance

Although this research does not aim to prescribe solutions, many interviews naturally turned toward the question of what can be done to make DAO spaces more inclusive, humane, and supportive. While no revolutionary ideas for a full-scale systemic overhaul were discovered, several contributors shared their small-scale practices they are already implementing or believe would be valuable to introduce. These suggestions do not rely on structural redesign, but instead focus on everyday interventions that challenge dominant norms and create space for cultural shifts from within. Following, some of the key narratives that surfaced will be discussed.

*(a) Education-first approach*

As highlighted by Veronica Kirin from Asterisk DAO, if DAOs genuinely want to welcome newcomers, especially those without prior Web3 experience, they should adopt an education-first approach. This means creating space for questions, exercising patience, and avoiding assumptions that everyone is on the same page. It also involves slowing down the pace of interactions, particularly in fast-moving Discord servers, where it's easy for people to feel overwhelmed or left behind. Normalising curiosity and not-knowing is important, and while this places extra work on more experienced contributors, it has a significant impact on inclusivity and retention. As shared by Veronica:

> We have to be ready to work with people who aren't necessarily that familiar with Web3. Make them feel comfortable even though we might be using terminology that they're not used to. Go a little bit slower. Explain the terminology.
>
> […]
>
> Basically every single time if they're not Web3 native, I have to have a one-on-one conversation to calm the nerves and to reinforce the fact that they are welcome and needed.

In parallel, there is a need for DAO contributors, particularly those in positions of influence, to develop awareness of the sociocultural dimensions of their power. In environments with no formal HR structures or accountability mechanisms, recognising how one's actions affect others becomes even more important. Building this awareness is by no means about enforcing hierarchy, but about cultivating responsibility in decentralised contexts.

*(b) Radical Honesty and Policy Building*

Some interviewees advocated for integrating values like *radical honesty*, conflict resolution protocols, and explicit anti-discrimination policies directly into DAO frameworks and not just as informal guidelines on Discord servers. Radical honesty, in this context, refers to a commitment to transparent communication and a willingness to address conflict directly. The benefit of this approach is in creating a culture where conflict is not avoided but expected and constructively engaged with for the sake of strengthened trust and accountability within the organisation.

As for policy building, there may be concerns that formalising inclusivity and anti-discrimination policies might replicate the bureaucratic hegemonic structures DAOs are positioned to overcome. Some interviewees, however, argued that structural clarity can in fact support decentralised trust.

*(c) Finding Accomplices*

A recurrently mentioned strategy was building informal alliances or, put simply, finding like-minded contributors to share concerns, and collectively resist toxic behaviours. This approach mirrors the dynamics of informal gatekeeping networks, but redirects them toward supporting inclusion and decentralisation. In spaces with no formal protection mechanisms, solidarity becomes a form of safety. Facing powerful actors alone can be intimidating, but having accomplices makes it easier to speak up and push back.

*(d) Calling Out Power Asymmetries and Discrimination*

Closely tied to the previous point, several contributors described the importance of actively calling out bias and discriminatory behaviour within DAOs. One interviewee shared how she *"regularly confront[s] whales or influential DAO figures"*, despite the social and economic risks involved. According to her, silence is not neutral, it only reinforces entrenched hierarchies. It is important to note that calling things out carries reputational risk and is emotionally difficult, particularly when it challenges the dominant culture of a space. This is precisely why finding accomplices is important, offering collective backing in moments that would otherwise feel isolating.

In the spirit of decentralisation, these reflections demonstrate that cultural change in DAOs tends to follow a bottom-up approach. It happens gradually through the building of relationships, creating supportive environments, and showing care for fellow contributors.

# 5. Conclusion

This report set out to explore the cultural and social dynamics across DAOs, spaces often celebrated for their decentralised ideals, yet marked by recurring patterns of exclusion and return to centralisation. Admittedly, a rather sobering picture has emerged through our qualitative inquiry: financial filters that favour the already privileged, reputation systems that reward visibility over substance, onboarding structures that presume prior fluency, and informal norms that quietly gatekeep participation. These findings may read as pessimistic, and perhaps they are. However, in this context, pessimism is not intended as defeatism. Instead, we see it as a form of grounded realism, as a refusal to romanticise decentralisation while systemic exclusions persist beneath the surface.

Importantly, this critique is not offered to oppose the idea of DAOs. We believe that DAOs do hold potential to reshape collective coordination and reimagine existing organisational governance structures. They are, still, one of the few arenas in the current digital landscape where technical infrastructure, community, and ownership are being negotiated in the open. This has to be acknowledged. The problems we have surfaced in this report are not the reasons to abandon the DAO experiments, but to strengthen it through deeper reflection and inclusive design.

For DAOs to realise their promise, they must move beyond the notion that decentralisation can be achieved through code or tokenomics alone. Social decentralisation is equally essential, and should be manifested through acts of recognition, reciprocal support, and intentional community-building. This report invites the reader to reflect on the dissonance between aspiration and reality, and to consider how more resilient and pluralistic DAO cultures might be created, beginning with one's own involvement.

# Acknowledgements

We thank Veronica Kirin, Sarah Smith, Natalie Crue, Gaia Soykok, Jose Martinez, RaidGuild members, Theodor Beutel, Joseph Low, Joshua Tan

# Bibliography

1. Ahmed, S. (2012). *On Being Included: Racism and Diversity in Institutional Life*. Duke University Press. https://doi.org/10.2307/j.ctv1131d2g

2. Austgen, J., Fábrega, A., Allen, S., Babel, K., Kelkar, M., & Juels, A. (2023). *DAO Decentralization: Voting-Bloc Entropy, Bribery, and Dark DAOs* (No. arXiv:2311.03530). arXiv. https://doi.org/10.48550/arXiv.2311.03530

3. Blommaert, J. (2018). *Durkheim and the Internet: On Sociolinguistics and the Sociological Imagination*. Bloomsbury Academic. https://doi.org/10.5040/9781350055223

4. Bourdieu, P. (1986). The forms of Capital. In *Handbook of Theory and Research for the Sociology of Education* (J. G. Richardson, pp. 241–258). Greenwood Press.

5. Bucher, T., & Helmond, A. (2018). The Affordances of Social Media Platforms. In J. Burgess, A. Marwick, & T. Poell, *The SAGE Handbook of Social Media* (pp. 233–253). SAGE Publications Ltd. https://doi.org/10.4135/9781473984066.n14

6. Crozier, M. (2010). *The bureaucratic phenomenon*. Transaction Publishers. https://doi.org/10.4324/9781315131092

7. De Filippi, P., & Merk, T. (2024). *How to DAO: The role of trust and confidence in institutional design* (SSRN Scholarly Paper No. 5071975). Social Science Research Network. https://doi.org/10.2139/ssrn.5071975

8. De Filippi, P., & Santolini, M. (2022). *Extitutional Theory: Modeling Structured Social Dynamics Beyond Institutions* (SSRN Scholarly Paper No. 4001721). Social Science Research Network. https://papers.ssrn.com/abstract=4001721

9. DuPont, Q. (2017). Experiments in algorithmic governance: A history and ethnography of "The DAO," a failed decentralized autonomous organization. In *Bitcoin and Beyond*. Routledge.

10. Dylan-Ennis, P. (2024). *Absolute Essentials of Ethereum*. Routledge. https://doi.org/10.4324/9781003319603

11. Dylan-Ennis, P., & Kavanagh, D. (2024). Hash, Bash, Cash—Decentralized Autonomous Organizations (DAOs) as a New Form of Democratic Organizing. In *Decentralized Autonomous Organizations*. Routledge.

12. Hassan, S., & De Filippi, P. (2021). Decentralized Autonomous Organization. *Internet Policy Review: Journal on Internet Regulation*, *10*(2), 1–10. https://doi.org/10.14763/2021.2.1556

13. Kerckhoven, S. V., & Chohan, U. W. (Eds). (2024). *Decentralized Autonomous Organizations: Innovation and Vulnerability in the Digital Economy*. Routledge. https://doi.org/10.4324/9781003449607

14. Merk, T. (2024, September 24). *The unusual DAO: An ethnography of building trust in "trustless" spaces* [Info:eu-repo/semantics/article]. Alexander von Humboldt Institute for Internet and Society gGmbH. https://doi.org/10.14763/2024.3.1795

15. Owocki, K. (with Puncar, & Tapscott, D.). (2025). *How to DAO: Mastering the Future of Internet Coordination* (1st ed). Penguin Publishing Group.

16. Pfaff-Czarnecka, J. (2011). From 'identity' to 'belonging' in social research: Plurality, social boundaries, and the politics of the self. In S. Albiez, N. Castro, L. Jüssen, & E. Youkhana (Eds), *Etnicidad, ciudadanía y pertenencia / Ethnicity, Citizenship and Belonging* (pp. 199–220). Iberoamericana Vervuert. https://doi.org/10.31819/9783954871124-010

17. Raj, P. M. (2022, September 20). 'DAOs are not Corporations': Ethereum's Vitalik Buterin. *Watcher Guru*.

    https://watcher.guru/news/daos-are-not-corporations-ethereums-vitalik-buterin

18. Sharma, T., Kwon, Y., Pongmala, K., Wang, H., Miller, A., Song, D., & Wang, Y. (2023). *Unpacking How Decentralized Autonomous Organizations (DAOs) Work in Practice* (No. arXiv:2304.09822). arXiv. https://doi.org/10.48550/arXiv.2304.09822

19. Tai, K. (2022, February 5). *DAOs are meant to be completely autonomous and decentralized, but are they?* Cointelegraph.

    https://cointelegraph.com/news/daos-are-meant-to-be-completely-autonomous-and-decentralized-but-are-they

20. Tan, J., Merk, T., Hubbard, S., Oak, E. R., Rong, H., Pirovich, J., Rennie, E., Hoefer, R., Zargham, M., Potts, J., Berg, C., Youngblom, R., Filippi, P. D., Frey, S., Strnad, J., Mannan, M., Nabben, K., Elrifai, S. N., Hartnell, J., … Boneh, D. (2024). *Open Problems in DAOs* (No. arXiv:2310.19201). arXiv. https://doi.org/10.48550/arXiv.2310.19201

21. Wajcman, J. (2004). *Technofeminism*. Polity.

22. Wright, A., & Law, C. P. of L. at B. N. C. S. of. (2021). The Rise of Decentralized Autonomous Organizations: Opportunities and Challenges. *Stanford Journal of Blockchain Law & Policy*. https://stanford-jblp.pubpub.org/pub/rise-of-daos/release/1

23. Zeilinger, M. (2023). Structures of belonging. *PostScriptUM*.